\documentstyle[epsfig,longtable]{aipproc}

\newcommand{\newc}{\newcommand}
\newcommand\eg{{\it {e.g.}}}
\newcommand\etal{{\it {et al.}}}
\newcommand\lsim{\mathrel{\rlap{\lower4pt\hbox{\hskip1pt$\sim$}}
    \raise1pt\hbox{$<$}}}
\newcommand\gsim{\mathrel{\rlap{\lower4pt\hbox{\hskip1pt$\sim$}}
    \raise1pt\hbox{$>$}}}

\newcommand\msusy{M_{SUSY}}
\newcommand\fa{f_{a}}
\newcommand\mchi{m_{\chi}}              

\newcommand\gravitino{\widetilde{G}}
\newcommand\mgravitino{m_{\gravitino}}

\newcommand\mhalf{m_{1/2}}      \newcommand\mzero{m_{0}}
\newcommand\mone{M_{1}}         \newcommand\mtwo{M_{2}} 

\newcommand\photino{\widetilde{\gamma}} \newcommand\mphotino{m_{\photino}}
\newcommand\gluino{\widetilde{g}} \newcommand\mgluino{m_{\gluino}}

\newcommand\axino{\widetilde{a}}        \newcommand\maxino{m_{\axino}}
\newcommand\abunda{\Omega_{\axino}h^2}

\newc{\cachigamma}{C_{a\chi\gamma}}

\newc{\caww}{C_{aWW}}                   \newc{\cayy}{C_{aYY}}
\newc{\sthw}{\sin\theta_W}              \newc{\cthw}{\cos\theta_W}
\newc{\bino}{\widetilde B}              \newc{\wino}{\widetilde W_3}
\newc{\higgsinob}{{\widetilde H}^0_b}   \newc{\higgsinot}{{\widetilde H}^0_t}

\newc{\abund}{\Omega h^2}
\newc{\abundchi}{\Omega_\chi h^2}
\newc{\rhocrit}{\rho_{crit}}
\newc{\rhochi}{\rho_{\chi}}

\newc{\mwimp}{m_{\rm WIMP}}     \newc{\rhowimp}{\rho_{\rm WIMP}}

\newc{\mplanck}{M_{\rm P}}              \newc{\mgut}{M_{\rm GUT}}
\newc{\mz}{m_{Z}}                       \newc{\mw}{m_{W}}

\newc{\xf}{x_f}
\newc{\jxf}{J({\xf})}
\newc{\VEV}[1]{\langle #1 \rangle}

\newcommand\tev{\,\mbox{TeV}}
\newcommand\gev{\,\mbox{GeV}}
\newcommand\mev{\,\mbox{MeV}}
\newcommand\kev{\,\mbox{keV}}
\newcommand\ev{\,\mbox{eV}}

\newc{\ra}{\rightarrow}
\newc{\beq}{\begin{equation}}
\newc{\eeq}{\end{equation}}
\newc{\bea}{\begin{eqnarray}}
\newc{\eea}{\end{eqnarray}}

\begin{document}
\begin{flushright}
{hep-ph/9903467\\
Spring Equinox, 1999}\\
\end{flushright}
\title{Non-Baryonic Dark Matter\\-- A Theoretical Perspective}

\author{Leszek Roszkowski$^*$\footnote{Invited review talk at
COSMO-98, the Second International Workshop on Particle Physics and
the Early Universe, Asilomar, USA,
November 15-20, 1998.}
}
\address{$^*$Department of Physics, Lancaster University, 
Lancaster LA1 4YB, England}

\maketitle

\begin{abstract}
I  review axions, neutralinos, axinos, gravitinos and super-massive
Wimpzillas as dark matter candidates.
\end{abstract}

\section*{Introduction}

Some twenty years after the dark matter (DM) problem was taken
seriously by particle theorists, we still don't know the exact nature
of the hypothetical non-luminous material of which presumably extended
halos around galaxies and their clusters are
made~\cite{dm-evidence}. While there may well be more than one type of
DM, arguments from large structures make us believe that a large, and
presumably dominant, fraction of DM in the Universe is made of some
massive particles which at the time of entering matter dominance would
be already non-relativistic, or {\em cold}.  From the particle physics
point of view, cold DM (CDM) could be made of some particles which
would generically be called weakly interacting massive particles
(WIMPs).

WIMPs do not necessarily have to interact only via weak interactions
{\em per se}.  One expects that they should preferably be electrically
and color neutral, and therefore be, as it is often stated,
``non-baryonic''. Otherwise, they would dissipate their kinetic
energy.  (Aspects of baryonic DM are discussed in
Ref.~\cite{griest-cosmo98} (MACHOs) and ~\cite{kusenko-cosmo98}
(Q-balls).)

Among WIMPs, there exist several interesting candidates for CDM which
are well-motivated by the underlying particle physics. The neutralino
of supersymmetry (SUSY) is considered by many a ``front-runner'' by
being perhaps the most ``generic'' WIMP.  The axion is another
well-motivated candidate. But by no means should one forget about
other possibilities. While some old picks (sneutrinos and neutrinos
with mass in the GeV range) are now ruled out, axinos and gravitinos
have recently been revamped as possibilities for CDM. A new type of
super-heavy Wimpzilla has also been proposed. In this talk I will
briefly summarize main results and review recent developments in the
field of WIMP and WIMP-like DM.

Neutrinos, the only WIMPs that are actually known to exist, are not
considered particularly attractive as DM candidates. It has long been
believed that their mass is probably very tiny, as suggested by
favoured solutions to the solar and atmospheric neutrino problems,
which would make them hot, rather than cold DM. This picture has
recently been given strong support by first direct evidence from
Superkamiokande for neutrinos' mass~\cite{superkamiokande}. While the
new data only gives the $\mu-\tau$ neutrino (mass)$^2$ difference of
$2.2\times10^{-3}\ev^2$, it it very
unlikely that there would exist two massive neutrinos with
cosmologically relevant mass of  5 to 40 eV and such a tiny mass
difference.

Current estimates of the lower bound on the age of the Universe lead to
$\abund<0.25$.  Recent results from high-redshift supernovae type Ia
imply $\Omega_{\rm matter}\simeq0.3$. The Hubble parameter is now
constrained to $0.65\pm0.1$. Since $\Omega_{\rm baryon} h^2\lsim0.015$,
one obtains $\Omega_{\rm CDM} h^2\lsim0.15$ or so. Assuming that CDM
accounts for most of matter in galactic halos, one obtains a very rough
estimate $\Omega_{\rm CDM} h^2\gsim0.025$.

\section*{Axions}

Axions are spin-zero particles which are predicted by the Peccei-Quinn
(PQ) solution~\cite{pq} to the strong CP problem. As it is well known,
the Lagrangian of QCD allows for a PC-violating term
${{\alpha_s}\over{8\pi}} \bar\theta G\widetilde G$, where $G$ is the
gluon field strength. This term would contribute about
$5\times10^{-16}\bar\theta\, {\rm e\,cm}$ to the electric dipole
moment of the neutron, thus violating the current experimental bound
by some ten orders of magnitude. In order to explain the required
strong suppression in the value of $\bar\theta$, Peccei and Quinn
postulated a new global $U(1)$ symmetry which would be spontaneously
broken at some scale $\fa$. The pseudogoldstone boson associated with
this scenario is the axion~\cite{axion}. Because of a QCD chiral
anomaly, the axion acquires mass $m_a\approx \Lambda_{QCD}^2/\fa$
where $\fa$ is {\it a priori} an arbitrary parameter.

Axions are also relevant cosmologically. A variety of astrophysical
and cosmological constraints have now narrowed the range of axion mass
to $10^{-6}\ev\lsim m_a\lsim 10^{-3}\ev$ (which corresponds to
$10^{(9-10)}\gev\lsim\fa\lsim10^{12}\gev$) in a broad range of axion
models. Rather remarkably, in this mass range, axion relic abundance
is of order one, making them a possible DM candidate. Because they
are produced out of thermal equilibrium, axions quickly become
non-relativistic and are cold relics. 

Axions are currently being searched for in microwave cavities immersed
in a strong magnetic field, as reviewed in a separate talk by
Sadoulet~\cite{sadoulet-cosmo98}.

\section*{Neutralinos}

The DM candidate which has attracted perhaps the most attention from
both the theoretical and experimental communities is the
neutralino. It is a neutral Majorana particle, the lightest of the
mass eigenstates of the fermionic partners of the gauge and Higgs
bosons: the bino, wino and the higgsinos. It is massive and, 
if it is the lightest SUSY
particle (LSP), it will be stable due to assumed R-parity. A
perfect candidate for a WIMP!  There is much literature devoted to the
neutralino as DM, including a number of excellent reviews (see, \eg,
Ref.~\cite{jkg}). Here I will only summarize the most essential
results and comment on recent developments and updates.

Neutralino properties as DM and ensuing implications for SUSY spectra
are quite model dependent but certain general conclusions can be
drawn. Two benchmark models are normally considered. One is the
supersymmetrized Standard Model (MSSM) with a minimum number of GUT
assumptions about the form of superpartner soft SUSY-breaking mass
terms. Of relevance here is the assumption that the masses of the
bino, wino and gluino (the fermionic partner of the gluon) are equal
at the GUT scale $\mgut\simeq2\times10^{16}\gev$.  The other, much
more predictive, model is the Constrained MSSM (CMSSM), or an
effective minimal supergravity model. In the CMSSM one assumes that
not only gaugino but also scalar (sfermion and Higgs) soft
SUSY-breaking masses unify to $\mhalf$ and $\mzero$, respectively, at
a GUT scale. Masses of all the superpartners at the Fermi scale are
then obtained by running RGE's down from $\mgut$ to $\mz$. Higgs
masses are determined through the condition of electroweak symmetry
breaking (EWSB).

It is now generally accepted that the neutralino as DM candidate
should most naturally contain a dominant bino
component~\cite{chiasdm}. The (admittedly rough and subjective
although commonly used) argument in the MSSM is that of naturalness:
no superpartner masses are expected to significantly exceed 1\tev.
This, because of GUT-related relations among the masses of the
gluino, the wino and the bino, $\mone\simeq 0.5\mtwo$ and
$\mtwo\simeq0.3\mgluino$, implies $\mchi\lsim150\gev$.
Of course this bound is only indicative but
it gives us some idea for the expected range of $\mchi$. Another
implication is that, because of the structure of the neutralino mass
matrix, higgsino-like neutralinos are also strongly disfavored (the
region $|\mu|\ll\mtwo$ where $\mu$ is the Higgs/higgsino mass
parameter)~\cite{chiasdm}.

Remarkably, in the CMSSM, the bino-like neutralino typically
automatically {\em comes out} to be the LSP in a very large part of
the SUSY parameter space~\cite{an93,kkrw}. This happens mostly as a
result of imposing EWSB which typically produces
$|\mu|\gg\mtwo$. Because of this property, and the fact that, roughly,
$\abundchi\sim \mzero^4/\mchi^2$, one is often able to put a {\em
cosmological} ($\abundchi<0.25$) upper bound on SUSY mass
parameters. Remarkably, the bound in the ball-park of 1\tev, as
generally expected for low-energy supersymmetry! To me this remarkable
property is a powerful illustration of the unity of particle physics and
cosmology.  While this attractive 
picture holds over large ranges of parameters,
it has some loop-holes. For large $\tan\beta$, the bound disappears in
a broad range around $m_A/2$ (half of the mass of the pseudoscalar
Higgs) where the annihilation cross-section becomes large. Another
effect which has recently been pointed out is the neutralino's
co-annihilation with the next lightest $\widetilde\tau_R$ in the
region of $\mhalf\gg\mzero$~\cite{ellis-cosmo98}.

Neutralinos and other SUSY particles have been constrained by searches
at LEP and elsewhere. Mass limits are somewhat model dependent but are
currently around $30\gev$ (MSSM) to $50\gev$ (CMSSM), as reviewed  by
J.~Ellis~\cite{ellis-cosmo98}. I also refer to his talk for an
explanation why higgsino-like LSPs are basically ruled out in a class
of CMSSM-like models (with somewhat relaxed assumptions about
unification of scalar masses). 

Detection of (SUSY) WIMPs follows two broad avenues, as reviewed in
separate talks~\cite{sadoulet-cosmo98,bergstrom-cosmo98}. 
Here I would like to make
some comments about possible evidence for a WIMP signal in annual
modulation. A superposition of the motion of the Earth around the Sun
with that of the Sun around the center of the Milky Way leads to a
small but sizeable (a few per cent) periodic variation in the
effective velocity of halo WIMPs and therefore also in the detection
rates~\cite{annualmodulation}. The rate should reach its peak on the
2nd of June. Based on 14,962\,${\rm day\times kg}$ of data, DAMA has
recently reported evidence of a possible signal in their NaI(Tl)
setup~\cite{dama98}, which also confirmed DAMA's earlier
indication~\cite{dama97} based on 4,549\,${\rm day\times kg}$ of data. DAMA
used a maximum likelihood method to compute in the $k$-th energy bin
the most probable value of $S_k=S_{0,k}+S_{t,k}\cos\left(\omega
(t-t_0)\right)$, where ($S_{0,k}$) $S_{t,k}$ are time (in-)dependent
components in the notation of Ref.~\cite{dama98}. The measured values
of $S_{0,k}$ and $S_{t,k}$ are fitted with two parameters
$\xi\sigma_p$ and $\mwimp$, where $\xi=\rhowimp/(0.3\gev/cm^3)$  and
$\rhowimp$ is the local density of WIMPs.

\begin{figure}[t!]      
\centerline{\epsfig{file=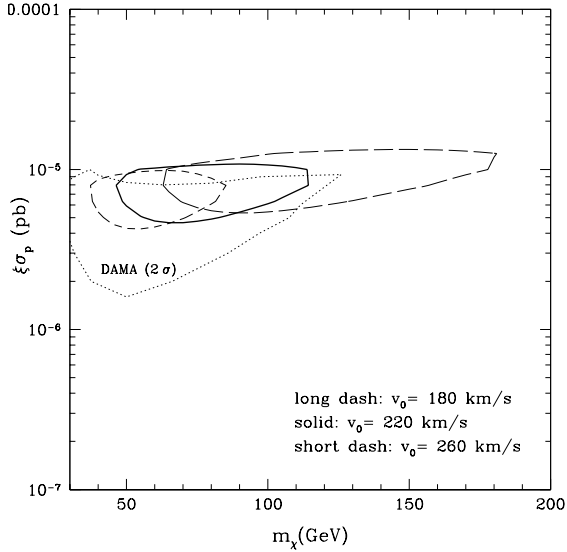,height=3in,width=3in}}
\bigskip
\caption{Contours of the function $\kappa=10$ defined in the text for
different values of the peak of the halo WIMP velocity
distribution. Denoted by dots is the $2\sigma$ region selected by DAMA
assuming $v_0=220\, km/s$. 
}
\label{kappa:fig}
\end{figure}

The derived ranges $\mwimp=59\gev^{+17}_{-14}\gev$ and
$\xi\sigma_p=7.0^{+0.4}_{-1.2}\times10^{-6}pb$ (at $99.6\%$ CL), are
in the ball-park of what one could expect from a genuine WIMP signal.
We should be looking with attention for other dedicated experiments
with similar sensitivity, like CDMS or UKDMC, to soon falsify the
effect. Here I would like to note that the region selected by DAMA is
probably too restrictive as has been shown in Ref.~\cite{michal}.
(See also Ref.~\cite{nfidm} for a crude estimate.) The effect is very
sensitive to assumptions about the form of the WIMP velocity
distribution in the halo.  In the analysis performed by DAMA only one
value of the peak of the Maxwellian velocity distribution  was
assumed,  $v_0=220\,
km/sec$, and only for the cored spherical isotermal model of the
halo. (Several other halo models were considered in Ref.~\cite{kk98}
and their effect on modifying direct detection rates was found to be
minimal.) 
Since the Galactic halo has not been directly
measured, quoted error bars for $v_0$ and the
local halo density should, in my opinion, be treated only as estimates
(if not ``guesstimates'').  Varying $v_0$ within a reasonable range
leads to a significant enlargement of the selected region. This can be
seen in Fig.~\ref{kappa:fig} taken from Ref.~\cite{michal} where we
plot the function $\kappa$ defined as 

\beq 
\kappa= \Sigma_{\rm energy\
bins} \left[ \left(S_{0,k}- S_{0,k}^{exp}\right)^2/\sigma_{0,k}^2
+\left(S_{t,k}- S_{t,k}^{exp}\right)^2/\sigma_{t,k}^2\right],
\label{kappa:eq}
\eeq 
where the $S_{.,.}$s and $S_{.,.}^{exp}$s are the calculated and
measured values and the $\sigma$'s are the experimental error bars
given in Table~2 of Ref.~\cite{dama98}. We minimize $\kappa$ to
determine $\xi\sigma_p$ and $\mwimp$ for a given value of $v_0$. One
can see in Fig.~\ref{kappa:fig} that the region of $\kappa<10$ which
broadly matches the $2\sigma$ region of DAMA strongly depends on
$v_0$. In particular, smaller values of $v_0$ allow for much larger
WIMP masses to be consistent with the possible signal from
DAMA~\cite{michal}.

Assuming that the effect reported by DAMA is caused by a genuine WIMP
signal, it is interesting to ask what ranges of SUSY parameters it
corresponds to. As reported here by Arnowitt~\cite{dickcosmo98talk}
and Fornengo~\cite{fornengo-cosmo98}, it is indeed
possible to find such SUSY configurations which could reproduce the
signal but only for large enough $\tan\beta\gsim10$. For smaller
$\tan\beta$ the mass of the pseudoscalar contributing to the
WIMP-nucleon cross-section becomes too small. The corresponding values
of $\abundchi$ are typically rather small, below 0.06 (MSSM) or 0.02
(CMSSM), although larger values can also be found. I expect these
conclusions to generally hold even if one neglects some arbitrariness in
enlarging the experimental region of DAMA~\cite{fornengo-cosmo98} (which I
don't find justified~\cite{michal-prep}), and in deciding which value of
{\em global} $\abundchi$ corresponds to the local WIMP
density~\cite{fornengo-cosmo98}. The effect of varying $v_0$ 
will have a significant impact on broadening the
experimental region and therefore also on the allowed configurations
of SUSY masses and couplings~\cite{michal-prep}.

\section*{Axinos}

Each of the two DM candidates described above resulted from an
attractive, albeit yet unconfirmed, idea in particle physics: the PQ
symmetry and SUSY. Taken together, these predict an axino, the
fermionic partner of the axion. Similarly to the axion, the axino
couples to ordinary matter with a very tiny coupling proportional to
$1/\fa$ where the allowed range of $\fa$s was given in discussing
axions. 

It is plausible to consider the axino as the LSP since its mass is
basically a free parameter which can only be determined in specific
models.  As we have seen above, the neutralino has been accepted in
the literature as a ``canonical'' candidate for the 
LSP and an attractive dark matter
candidate. But with current LEP bounds between 30 and 50~GeV, 
it becomes increasingly plausible that there
may well be another SUSY particle which will be lighter than the
neutralino, and therefore a candidate for the LSP and dark matter.

Primordial axinos decouple from the thermal soup very early, around
$T\simeq\fa$, similarly to the axions. The early study of Ragagopal,
Turner and Wilczek~\cite{rtw} concluded that, in order to satisfy
$\Omega h^2<1$, the primordial axino had to be light ($\lsim 2$ keV),
corresponding to warm dark matter, unless inflation would be invoked
to dilute their abundance. In either case, one did not end up with
axino as cold DM.

\begin{figure}[t!]      
\centerline{\psfig{file=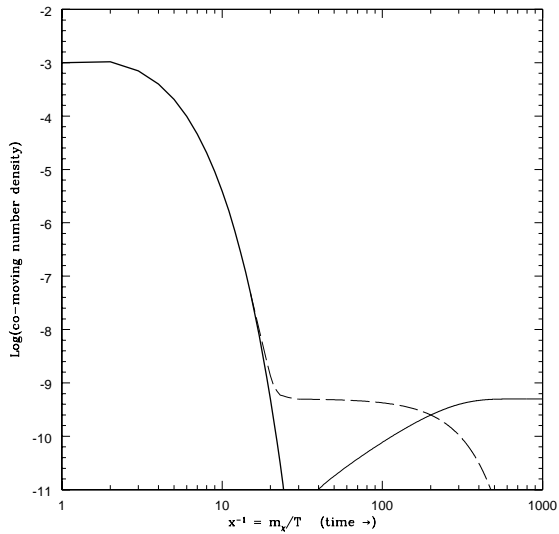,height=3in,width=3in}}
\bigskip
\caption{
A schematic behavior of the co-moving number density: the thermal
equilibrium (thick solid), NLSP neutralino (dash) and LSP axino (thin
solid). 
}
\label{freezeout:fig}
\end{figure}

It has recently been shown~\cite{ckr} that the axino can be a
plausible {\em cold} dark matter candidate after all, and that its
relic density can naturally be of order the critical density. The
axino can be produced as a non-thermal relic in the decays of heavier
SUSY particles. Because its coupling is so much weaker, superparticles
first cascade decay to the next lightest SUSY partner (NLSP) for which
the most natural candidate would be the neutralino. The neutralino
then freezes out from thermal equilibrium at $T_f\simeq\mchi/20$. If
it were the LSP, its co-moving number density after freeze-out would
remain nearly constant. In the scenario of~\cite{ckr}, the
neutralino, after decoupling from the thermal equilibrium, 
subsequently decays into the axino via, \eg, the process
\begin{equation}
\chi\ra\axino\gamma
\label{chitoagamma:eq}
\end{equation}
as shown in Fig.~\ref{freezeout:fig}. This process was already
considered early on in Ref.~\cite{kmn} (see also~\cite{rtw}) in the
limit of a photino NLSP and only for both the photino and axino masses
assumed to be very low, $\mphotino \leq 1 \gev$ and $\maxino\leq 300
\ev$, the former case now excluded by experiment. In that case, the
photino lifetime was typically much larger than 1 second thus normally
causing destruction of primordial deuterium from Big Bang
nucleosynthesis (BBN) by the energetic photon of
(\ref{chitoagamma:eq}). 
Avoiding this led to lower
bounds on the mass of the photino, as a function of $\fa$, in the
$\mev$ range~\cite{kmn}.

Because both the NLSP neutralino and the CKR axino are both 
heavy (GeV mass range), the decay~(\ref{chitoagamma:eq}) is now typically
very fast. 
In the theoretically most favored case of a
nearly pure bino~\cite{chiasdm,kkrw}, 
the neutralino lifetime can be written as
\beq
\tau\simeq 2.12 \times 10^{-1} {\rm sec}\, \left({1\over{N\cayy}}\right)^2
\left(\frac{f_a}{10^{11}\gev}\right)^2
\left(\frac{50\gev}{\mchi}\right)^3
\label{binolife:eq}
\eeq
where the factor $N\cayy$ is of order one.
One can see that it is not difficult to ensure that the decay takes
place well before 1 second in order for avoid 
problems with destroying successful predictions of Big Bang
nucleosynthesis.  The axino number density is equal to that of the
NLSP neutralino. Therefore its relic abundance is
\beq
\abunda= \left({\maxino\over\mchi}\right)\abundchi.
\label{abunda:eq}
\eeq
The axinos are initially relativistic but, by the time of matter dominance
they become
redshifted by the expansion and become cold DM.

\section*{Gravitinos}

Another old DM candidate has recently been re-analyzed. In the context of
supergravity, there exists the fermionic partner of the graviton. Just
like the graviton, it couples to ordinary matter only gravitationally
with the strength $1/\mplanck$ where $\mplanck=2.4\times10^{18}\gev$.
Gravitino's mass is model dependent but is expected to be of order the
SUSY breaking scale $\msusy\lsim1\tev$. 

The story of primordially produced gravitinos is analogous to that
of axinos. Along with gravitons, they decouple at $T\sim\mplanck$. 
If they were the LSPs, they would ``overclose the
Universe'', unless either $\mgravitino\lsim2\kev$ or inflation
followed to dilute their density. In order for subsequent reheating
not to re-populate them, one requires $T_{\rm reh}\lsim10^9\gev$. 

Recently, a scenario for producing thermal gravitinos in the context
of leptogenesis has been considered~\cite{bbp98}. Decays of heavy
($\sim10^{16}\gev$) Majorana neutrinos violate $L$ which, via
$B-L$, leads to baryon asymmetry consistent with the observed values
$n_B/s\sim10^{-9}$. For this picture to work, and assuming
hierarchical neutrino masses, the baryogenesis temperature of
$T_B\sim10^{10}\gev$, and therefore  at least as large $T_{\rm reh}$,
are required. Gravitinos are then produced in the thermal bath
mainly via two-body processes involving gluinos. Their resulting relic
abundance is given by~\cite{bbp98} 
\beq
\Omega_{\gravitino}h^2= 0.60\left({T_B}\over{10^{10}\gev}\right)
\left({100\gev}\over{\mgravitino}\right)
\left({\mgluino(\mu)}\over{1\tev}\right)^2
\label{gravitinoabund:eq}
\eeq
where $\mu\sim100\gev$. One can see that for plausible values of $T_B,
\mgravitino$ and $\mgluino$ one obtains $\Omega_{\gravitino}h^2\sim1$.

One still has to make sure that the gravitinos are not produced in large
numbers in out-of-equilibrium decays of the NLSP which, because of
gravitino's tiny couplings, would take place long after BBN. There are
ways to satisfy this, for example, when the NLSP is a higgsino in the
mass range between $m_W$ and some $300\gev$ for which 
$\Omega h^2$ is small enough ($<0.008$). We note here that this
requirement should be easily satisfied also for a bino-like NLSPs if
one, or more, of the scalar leptons and squarks (except perhaps for sneutrinos
and stops) is sufficiently light to suppress bino's relic abundance.

\section*{Super-heavy WIMPs: Wimpzillas}

We have seen that plausible WIMP candidates need not couple to
ordinary matter only via weak interactions, nor do
they have to be produced thermally (axinos). Other possible candidates
exist. For example, in a stringy scenario, these may be some class
of moduli~\cite{brustein-cosmo98}. An interesting class of
non-thermally produced  
super-heavy relics
has recently been suggested~\cite{kolb-cosmo98}. (Thermally-produced
WIMPs of mass above some $500\tev$ would give $\Omega
h^2>1$~\cite{gk90}.)  Dubbed Wimpzillas, such relics could be as heavy
as $M_{\rm GUT}$. Moreover, they could even carry electric or color
charge.  One has to make sure that Wimpzillas do not annihilate
efficiently enough to  be in chemical equilibrium at any time. This is
basically guaranteed because their number density must be very tiny in
order not to ``overclose'' the Universe~\cite{kolb-cosmo98}.

Several possible mechanisms for generating Wimpzilla-like candidates
have been suggested. They could for example be
produced as a result of ``freezing out''  quantum fluctuations at
the end of inflation or by gravitational effects. First-hand
description of several of them can be found in these
Proceedings~\cite{kolb-cosmo98,tkachev-cosmo98}. 

It is remarkable that the Universe, and our halo, may well be filled
with such obese relics which, despite possibly large couplings to
ordinary matter, would be gentle enough to remain in the (dark)
background. At the end, the frightening Wimpzilla may reveal itself as
a ``monster with a human face''.

\section*{Summary}
Who is the WIMP? The key to answering this question will ultimately be
in the hands of experimentalists. Some attractive candidates (axion,
neutralino) will hopefully be either discovered or basically ruled out during
the next decade.  
Axinos and gravitinos, and
perhaps also Wimpzillas, may well have to wait for future generations
of dark matter enthusiasts.

\section*{Acknowledgements}
I am greatly indebted to David Caldwell and other members of the Local
Organizing Committee for setting up an inspiring meeting in one of the
most beautiful corners in the world, in the spirit of COSMO workshops.


\begin{references}

\bibitem{dm-evidence} See, \eg, M.S.~Turner, in the Proceedings.

\bibitem{griest-cosmo98} K.~Griest,  in the Proceedings.
\bibitem{kusenko-cosmo98} A.~Kusenko,  in the Proceedings.

\bibitem{superkamiokande} Super-Kamiokande Collaboration, talk by
Y. Suzuki at {\it Neutrino-98}, Takayama, Japan, June 1998; 
Super-Kamiokande Collaboration, Y.~Fukuda, \etal,
Phys. Lett. {\bf B433}, 9 (1998);
Phys. Lett. {\bf B436}, 33 (1998);
Phys. Rev. Lett. {\bf 81}, 1562 (1998).

\bibitem{pq} R. D.~Peccei and H. R.~Quinn, Phys. Rev. Lett. {\bf 38},
1440 (1977); Phys. Rev. {\bf D16}, 1791 (1977).
\bibitem{axion} S.~Weinberg, Phys. Rev. Lett. {\bf 40}, 223 (1978);
F.~Wilczek, Phys. Rev. Lett. {\bf 40}, 279 (1978).

\bibitem{sadoulet-cosmo98} B.~Sadoulet, in the Proceedings.

\bibitem{jkg} G.~Jungman, M.~Kamionkowski, K.~Griest, Phys. Rep. {\bf
267},  195 (1996). 

\bibitem{chiasdm} L.~Roszkowski, Phys. Lett. {\bf B 262}, 59 (1991);
see also J.~Ellis, D.V.~Nanopoulos, L.~Roszkowski, and D.N.~Schramm, 
Phys. Lett. {\bf B245}, 251 (1990).

\bibitem{an93} P.~Nath and R.~Arnowitt, Phys. Lett. 
{\bf B289}, 368 (1992).

\bibitem{kkrw} R.G.~Roberts and L.~Roszkowski, Phys. Lett. 
{\bf B309}, 329 (1993); 
G.L.~Kane, C.~Kolda, L.~Roszkowski, and J.~Wells, 
Phys. Rev. {\bf D49},  6173 (1994).

\bibitem{ellis-cosmo98} J.~Ellis, in the Proceedings and
references therein.
\bibitem{bergstrom-cosmo98} L.~Bergstr{\"o}m, in the Proceedings.

\bibitem{annualmodulation} A.~Drukier, \etal, Phys. Rev. {\bf D33},
3495 (1986); K.~Freese, \etal, Phys. Rev. {\bf D37}, 3388 (1988).
\bibitem{dama98} R.~Bernabei, \etal\ (The DAMA Collaboration),
ROM2F/98/34 (August 1998).
\bibitem{dama97} R.~Bernabei, \etal\ (The DAMA Collaboration), Phys. Lett. 
{\bf B424}, 195 (1998).

\bibitem{michal} M.~Brhlik and L.~Roszkowski, hep-ph/9903468.

\bibitem{kk98} M.~Kamionkowski and A.~Kinkhabwala, Phys. Rev. {\bf
D57} 3256 (1998).

\bibitem{michal-prep} M.~Brhlik and L.~Roszkowski, to be published.

\bibitem{nfidm} N.~Fornengo, talk at IDM-98, Buxton, UK, September
'98, hep-ph/9812210.
\bibitem{dickcosmo98talk} R.~Arnowitt,  in the Proceedings;
R.~Arnowitt and P.~Nath, hep-ph/9902237.
\bibitem{fornengo-cosmo98} N.~Fornengo,  in the Proceedings and
references therein to A.~Bottino, \etal, hep-ph/9808456,
hep-ph/9808459, and hep-ph/9809239.

\bibitem{rtw}
K.~Ragagopal, M.S.~Turner, and F.~Wilczek, Nucl. Phys. {\bf B358},
447 (1991).
\bibitem{ckr} L.~Covi, L.~Roszkowski and J.E.~Kim, LANCS-TH/9824
(December 1998), submitted to Phys. Rev. Lett.
\bibitem{kmn} J.E.~Kim, A. Masiero, and D.V.~Nanopoulos, Phys. Lett.
{\bf B139}, 346 (1984).

\bibitem{bbp98} M. Bolz, W. Buchm{\"u}ller, and M. Pl{\"u}macher, 
Phys. Lett. {\bf B443}, 209 (1998).

\bibitem{brustein-cosmo98} R.~Brustein,  in the Proceedings;
R.~Brustein and M.~Hadad, hep-ph/9810526.

\bibitem{kolb-cosmo98} E.~Kolb, in the Proceedings; D.J.H.~Chung,
E.W.~Kolb, and A.~Riotto, Phys. Rev. Lett. {\bf 81}, 4048 (1998). 

\bibitem{gk90} K.~Griest and M.~Kamionkowski, Phys. Rev. Lett. {\bf
64}, 615 (1990).

\bibitem{tkachev-cosmo98} I.~Tkachev, in the Proceedings; V.~Kuzmin
and I.~Tkachev, hep-ph/9809547.




\end{references}
\end{document}